# Tunable and Sensitive Detection of Cortisol using Anisotropic Phosphorene with a Surface Plasmon Resonance Technique: Numerical Investigation

Vipin Kumar Verma,[a,c] Sarika Pal,[a] Conrad Rizal[b], Yogendra Kumar Prajapati[d*]

[a]National Institute of Technology Uttarakhand, ECE Department, Srinagar, Pauri Garhwal, India, 246174
[b]Seed NanoTech International Inc., Brampton, ON, L6Y 3J6 Canada
[c]KIET Group of Institutions, Delhi-NCR Ghaziabad, 201206
[d]Motilal Nehru National Institute of Technology Allahabad, ECE Department, Prayagraj, India, 211004

**Abstract**. Tunable and ultrasensitive surface plasmon resonance (SPR) sensors are highly desirable for monitoring stress hormones such as cortisol, a steroid hormone formed in the human body's adrenal glands. This paper describes the detection of cortisol using a bimetallic SPR sensor based on highly anisotropic two-dimensional material, *i.e.*, phosphorene. Thicknesses of bi-metal layers, such as copper (Cu) and nickel (Ni), is optimized to achieve strong SPR excitation. The proposed sensor is rotated in-plane with a rotation angle (φ) around the z-axis to obtain phosphorene anisotropic behavior. The performance parameters of the sensor are demonstrated in terms of higher sensitivity (347.78 $^{o}$/RIU), maximum angular figure of merit (FOM*= 1780.3), and finer limit of detection (0.026 ng/ml). Furthermore, a significant penetration depth (203 nm) is achieved for the proposed sensor. The obtained results of the above parameters indicate that the proposed sensor outperforms the previously reported papers in the literature on cortisol detection using the SPR technique.

***Corresponding Author E-mail: yogendrapra@mnnit.ac.in

## 1 Introduction

Cortisol is a steroid hormone and is commonly referred to as the stress hormone because of its connection to the stress response of human beings. The cortisol level in blood and saliva impacts the cardiovascular processes, blood pressure, and many other metabolic activities [1]. A rise in cortisol level may cause Cushing syndrome - a fatty hump between shoulders, a rounded face, and various stretch marks on the skin. In contrast, cortisol insufficiency can cause Addison disease - an uncommon disorder that occurs when the body cannot produce enough of certain hormones [1]. Measuring cortisol levels is necessary for determining its deficiency, saturation levels and identifying various disorders linked to it. Cortisol levels can be determined using a variety of body fluids, including blood (invasive) and saliva (non-invasive) [1-2]. Its status in the saliva is straightforward to measure because these samples keep their original properties for at least a week or more [1]. Chemiresistor, electrochemical sensor, impedimetric biosensor, and surface plasmon resonance (SPR) sensors can all be used to measure or quantify cortisol levels [3-7]. SPR sensors are more suited than other mentioned sensors due to their quick, real-time, label-free sensing capabilities [8]. The SPR sensor is a powerful tool that detects molecular interactions by changing the probing medium's refractive index (RI).

To study the level of cortisol in saliva, Steven et al. reported a six-channel portable SPR biosensor based on competition assays [1]. The detection limit (0.36 ng/mL) of cortisol in laboratory buffers was determined. Further, to detect cortisol levels, researchers proposed and used gold (Au) nanoparticle-based long-range surface plasmon resonance (LRSPR) sensors, grating-based fiber optic sensors, and lossy mode resonance-based SPR sensors [8-12]. Some of the critical experimental works for cortisol detection are reported in the literature based on techniques such as electrochemical immuno-sensor, aptamer-based Au nanoparticles, and SPR immunoassay. However, the limit of detection (LOD) values using these techniques are still lower ($\leq$ 1ng/ml), respectively [11-13].

Recently researchers demonstrated the improved performance of the SPR sensor by utilizing the two-dimensional (2-D) materials such as graphene, transition metal di-chalcogenides (TMDs), few layers of phosphorene - a monolayer of black phosphorus (BP), antimonene, and $Ti_3C_2T_x$-Mxene over a metal layer due to their high charge carrier mobility, large work function, high adsorption energy, large surface area, stronger interaction with biomolecules, and high chemical stability [14-18]. Therefore, 2-D materials have a high potential for sensing, photonic, optoelectronic applications, and also their use in SPR sensors for cortisol detection remains elusive. However, the zero bandgaps of graphene, low carrier mobility, and hydrophobicity of TMDs, the narrow energy bandgap of MXene became their fatal disadvantages [19]. Alternatively, phosphorene is a stable 2-D layered material possessing excellent hole mobility (10,000 $cm^2V^{-1}s^{-1}$), tunable bandgap (0.3-2 eV), attention-grabbing puckered surface morphology, strong binding energy, hydrophilic nature, 40 times higher molar response factor (even more significant than graphene and TMDs), and parts per billion (ppb) sensing ability and have shown great potential for gas, humidity, and biosensing [20-24]. Thus, phosphorene's sensitivity, S, and selectivity for water vapors and gas can be tuned to advantage when used for sensing cortisol concentration in saliva. However, long-time direct exposure of phosphorene to the ambient environment may lead to its oxidation [18]. So, care must be taken during the experimental process using a high-quality and stable phosphorene sheet to get accurate results.

The most exceptional property of phosphorene is its in-plane anisotropy arising from its $sp^3$ hybridized puckered lattice structure [25]. Utilizing in-plane anisotropy feature of phosphorene to produce a tunable sensor device, this work aims to enhance the sensor performance in terms of sensitivity and LOD for detecting cortisol concentration at an operating wavelength, $\lambda$ = 830 nm [8]. However, for the tunability aspect, the strongest excitation of plasmons on the metal-phosphorene interface is considered through the rotation angle ($\varphi$) of the integrated device around the $z$-axis in-plane, resulting in variable charge transfer between phosphorene and the metal, leading to change in minimum reflectivity. In a broader sense, the objective of this paper is to demonstrate tunable sensitivity by simply rotating the integrated device around the z-axis in-plane. More importantly, the uniqueness of this work is two-prone. First, we have proposed a sensor for cortisol concentration detection with enhanced performance parameters (sensitivity, LOD, etc.) over the existing state-of-the-art. Second, we have introduced and analyzed the influence of the anisotropic behavior of phosphorene to enhance the charge transfer between phosphorene and the metal layer, which results in the finest angular figure of merit (FOM*) over the other existing cortisol concentration sensors.

## 2 Theoretical Modeling, Performance Parameters and Experimental Feasibility

Fig. 1 shows the proposed Kretschmann configured SPR sensor for cortisol concentration sensing. It consists of BK-7 prism, bi-metal layers (Cu/Ni), phosphorene, and cortisol saliva solutions. The working principle of the proposed sensor is based on surface plasmon resonance conditions. Surface plasmons (SPs) are quanta of charge density oscillations at the metal/dielectric interface that are excited by incident p-polarized light with coupling through a prism, grating, or waveguide [8], [15-17], [26]. Prism coupling is a preferred technique for its realizable and straightforward geometry [8, 16], where the resonance condition can be achieved after matching of wave vector of incident light with the wave vector of surface plasmon wave (SPW) ($k_{inc.} = k_{spw}$). This matching condition is susceptible to changes in the refractive index (RI) of the probing media due to adsorption of analyte on the sensor surface, which may be exploited for imaging and sensing applications [16]. Here, a low RI prism is chosen for enhanced light coupling [27]. The two combinations of bi-metal layers are used to identify the best metal layers for enhanced sensitivity. Salivary solutions containing different cortisol concentrations are considered an analyte medium where the RI values at different cortisol concentrations are taken from R.C. Stevens *et al.* [1]. The RIs and optimized thicknesses of all constituent layers are presented in Table 1. The RIs of all isotropic layers are fixed except for phosphorene, whose RI is directional-dependent due to its anisotropic nature. Phosphorene - a monolayer of black phosphorus (BP) is an orthorhombic layered crystal with two optical axes: $C_1$ and $C_2$, whose biaxial dielectric coordinate system does not lie parallel to the crystal coordinate subsystem [25]. In order to obtain the strongest SPR excitation, a well-matched condition between two optical axes of phosphorene and SPR wave is needed. One way to achieve this is by tuning the rotation angle, $\varphi$ of the SPR device around the z-axis in the plane. Once the angle $\varphi$ is fixed, incident angle θ is varied from 1 to $90^{o,}$ and RI of phosphorene is calculated using transfer matrix method (TMM) modeling proposed in Ref. [25]. The TMM is an accurate and popular method used to calculate the reflected intensity of p-polarized incident light [14,15]. The TMM for the reflectance calculation of the proposed sensor has been discussed in the supplementary material (SM1). In the present case, we have used MATLAB software to simulate the results (*i.e.*, to obtain reflectance curves) and analyze the sensor performance analytically in terms of sensitivity ($S = \frac{\Delta\theta_{SPR}}{\Delta n_a}$ [°/RIU]), minimum reflectance ($R_{min}$), full width at half maximum (FWHM), detection accuracy (DA=1/FWHM), the limit of detection (LOD=$\frac{\Delta C_{Conc.}}{\Delta\theta_{SPR}} \times 0.001^o$ [ng/mL]), figure of merit (FOM= $S \times DA$ [$RIU^{-1}$]), and angular figure of merit (FOM*= $\left|\frac{dR(\theta)/dn_a}{R(\theta)}\right|$) [25-26], here, * indicates the angular word. The angular detection limit of $0.001^o$ is taken for the angular interrogated SPR sensor [26]. Note that FWHM measures the angular width of the SPR curve at 50% reflection intensity, *i.e.*, $R_{min}$= 0.5a.u.

Table 1. Proposed sensor's thickness and RIs of constituent layers at the characteristic wavelength at λ = 830nm

| **Constituent Layers** | **Thickness (nm)** | **Refractive Index (RI)** |
|---|---|---|
| BK-7 Prism | - | 1.5102 [18] |
| Copper layer | 15 | 0.10807+$i$× 5.3990 [27] |
| Nickel layer | 80 | 2.2777+$i$× 5.0030 [27] |
| Phosphorene (BP) | 6×0.53 | Optimized by tuning with the rotation angle ($\varphi$) of the sensor around the $z$-axis [25] |

**2.1 Fabrication and Experimentation Feasibility of Proposed Sensor:**

The following steps should be taken to prepare a sensor chip for cortisol concentration sensing. First, the Cu layer can be grown on the BK-7 prism using electron beam evaporation or similar high vacuum deposition techniques [27-28]. Fabrication of Ni layer over Cu in the same fabrication cycle can be carried out to avoid oxidation of Cu layer. After that, the phosphorene layer, which may be obtained through the liquid phase exfoliation technique, can be chemically transferred over the Ni layer [27-28]. For cortisol sensing, a specially designed in-line filtering flow cell containing salivary solution with different cortisol concentrations can be used to deliver the small size organic cortisol molecule over the phosphorene surface of the sensor chip [1]. Also, the use of specially designed in-line filtering flow cells will be helpful to reduce non-specific binding by retaining the larger size biomolecules in salivary solution [1]. The interaction of cortisol molecule with phosphorene surface leads to physio adsorption of cortisol molecule on phosphorene surface through weak hydrogen bonding between phosphorus atom of phosphorene and a hydrogen atom bonded with the oxygen atom of cortisol molecule [29]. Stronger binding energy, the puckered surface morphology of phosphorene, and the lone pair of electrons of each phosphorus atom can be effectively utilized to facilitate the adsorption of cortisol molecule on sensor surface [15], [22-24], [29]. This molecular interaction leads to a RI shift of the sensing medium, which may be detected by measuring the resonance angle shift of the reflectance curve. To calculate the resonance angle shift via experimental investigation, the complete combination of prepared sensor chips placed over the one of the flat faces of the BK-7 prism may be put over the rotary base of the goniometer. Then, a monochromatic light source (AlGaAs laser) operating at λ = 830 nm may be used to illuminate the SPR chip via prism [8]. The resonance angle can be set by rotating the rotary base of the goniometer with the help of computer-controlled software. Finally, reflected light from the prism at a different incident angle may be measured through a photodetector and record the resonance angle shift after binding of cortisol molecule on the sensor surface.

Further, a more sensitive and selective attachment of cortisol molecules on the sensor surface, a suitable cortisol-sensitive affinity layer (anti-cortisol antibodies via cysteamine), can also be used [30]. In this work, we used experimentally measured RI values ranging from 1.3297 to 1.3311, corresponding to different cortisol concentrations adapted from Ref. [1] for numerical simulations. However, experimentally obtained results may vary slightly from the analytical simulations presented here as there are several unavoidable parameters during experimental realization that may affect the sensor performance.

# 3 Results & Discussion

## *3.1 Metal Thickness Optimization and Reflectance Curves*

Previous studies reveal that the metal layer plays a significant role in the generation of surface plasmons (SPs) and the attachment of analytes during SPR sensing applications [25], [27]. However, the use of single metal shows less sensitivity due to poor attachment of analyte on it, which can be further enhanced by using bi-metal layers. According to the literature, using a bi-metal layer combination in the SPR sensor results in higher sensitivity and accuracy [27]. Moreover, optimizing metal layer thicknesses is essential to achieve the nearly minimum reflectance ($R_{min}$) value at the resonance angle for attaining high DA and FOM*. Therefore, we have verified the sensor performance in maximal sensitivity and $R_{min}$ for a few Cu/Ni metal layer thickness combinations. For example, the metal thickness combination of Cu/Ni is 40/35 nm, meaning that the thickness of Cu and Ni is 40 nm and 35 nm, respectively.

Fig. 2 shows the optimization of Cu/Ni thicknesses at monolayer anisotropic phosphorene in terms of sensitivity and $R_{min}$ with the balance of photon absorption energy and electron energy loss [25]. Propagation of SPR waves in phosphorene, unlike isotropic 2D nanomaterial (e.g., graphene, TMDs), is significantly different [22]. Hence, the strongest SPR excitation can be achieved only after perfect matching optical axes of phosphorene crystal with p-polarized incident light by tuning the rotation angle ($\varphi$) [25]. In this regard, the $\varphi$ is optimized for each thickness combination of Cu/Ni by rotating the proposed sensor around the z-axis. Fig. 2 plotted at the optimized value of $\varphi$, *i.e.*, 72$^{o,}$ depicts the highest sensitivity of 229.18 $^{o}$/RIU at smaller $R_{min}$= 0.0612 for the thickness of Cu (15nm), Ni (80nm), and monolayer BP. Further, the influence of $\varphi$ on performance for the proposed sensor has been discussed in the supplementary material (SM2).

In contrast, sensitivity lies below this. for all other thickness combinations of Cu and Ni. The optical anisotropy of BP can be effectively tuned by adjusting the number of BP layers, as the optical conductivity of BP layers changes by varying its thickness [18], [20-21]. After optimizing the thickness combination of bimetal layers of the proposed sensor, we evaluated the sensor performance parameters such as S, DA, and figure of merit (FOM) for 0 to 6 layers of BP, as shown in Table 2. It may be analyzed from Table 2 that all performance parameters enhance increasing the number of BP layers. For example, the maximum sensitivity (320.86 °/RIU), DA (0.7042 1/°), and FOM (225.96 $RIU^{-1}$) are obtained for six layers of BP. The improved performance is due to light-matter solid interaction observed at higher BP layers because of more significant adsorption energy and optical conductivity.

Further, the increment of BP layers shows higher $R_{min}$ for the proposed sensor. So, the parameters are evaluated only for 0-6 layers of BP. Therefore, all subsequent simulations for cortisol sensing are performed at optimized thicknesses of Cu (15nm), Ni (80nm), and six layers of BP.

Table 2: Sensor parameters evaluated for 0-6 layers of BP

| BP layers | $\theta_{res.}$ at $n_s$=1.33 (°) | $\theta_{res.}$ at $n_s$= 1.3305 (°) | $R_{min}$ (a.u.) | S (°/RIU) | FWHM (°) | DA (1/°) | FOM ($RIU^{-1}$) |
|---|---|---|---|---|---|---|---|
| 1 | 76.7935 | 76.6789 | 0.0612 | 229.18 | 1.89 | 0.5291 | 121.26 |
| 2 | 77.0972 | 76.9826 | 0.0580 | 240.56 | 1.87 | 0.5348 | 128.64 |
| 3 | 77.4353 | 77.3092 | 0.0506 | 252.10 | 1.80 | 0.5556 | 140.06 |
| 4 | 77.8134 | 77.6816 | 0.0417 | 263.56 | 1.78 | 0.5618 | 148.07 |
| 5 | 78.2374 | 78.0942 | 0.0037 | 286.48 | 1.69 | 0.5917 | 169.51 |
| 6 | 78.7157 | 78.5553 | 0.0252 | 320.86 | 1.42 | 0.7042 | 225.96 |

The reflectance curve is simulated for different cortisol concentrations, *i.e.*, 0.36, 0.72, 1.80, 3.60, and 4.50 ng/ml at optimized Cu (15nm) and Ni (80nm) thicknesses, and six layers of anisotropic phosphorene.

Fig. 3 shows the reflectance curve at various cortisol concentrations (0.36 to 4.6 ng/ml). As shown,

the resonance angle shifts to a higher value, and $R_{min}$ also gets more significant for higher cortisol concentrations. The improvement is due to higher binding energy, and larger surface area offered due to puckered surface morphology of phosphorene for efficient adsorption of cortisol present in saliva [21], [25]. The variation in RI of the analyte medium with adsorption of cortisol modifies the matching condition of wave vector of evanescent field and SPW at some other higher incident angle.

Fig. 4 illustrates the variation of SPR angle, $R_{min}$ and FWHM estimated from the reflectance curves in Fig. 3. The increase in SPR angle and larger values of $R_{min}$ for higher cortisol concentrations indicates the sensitive detection of cortisol concentration in saliva. The smaller FWHM is observed at higher cortisol concentrations due to the minimum damping of SPs [25]. Likewise, as shown in Fig.5, cortisol concentrations versus sensitivity and FOM are also evaluated using reflectance curves in Fig. 3. The cortisol concentration of 0.36 ng/ml is used as a reference sample for sensitivity and FOM calculations. The rising trend in sensitivity and FOM is related to the sensitive and accurate detection of cortisol concentrations in saliva. A precise observation from Fig. 5 indicates that the maximum sensitivity of 343.78 $^{o}$/RIU and FOM of 1780.3 $RIU^{-1}$ are observed at 4.5 ng/ml of cortisol concentration.

### *3.2 Analysis of FOM* and LOD*

LOD and FOM* are two critical parameters to indicate the most exemplary performance of a proposed sensor for cortisol concentration sensing. The highest position of the FOM* curve slope tells the maximum value of FOM* [25]. LOD signifies the lowest possible cortisol concentration that can be detected. In summary, the maximum FOM* and the smallest possible LOD value are desired to enhance cortisol concentration sensing [25].

FOM* curve at various cortisol concentrations is depicted in Fig. 6. As shown in it, the maximum FOM* values obtained for cortisol concentrations of 0.36, 0.72, 1.80, 3.60, and 4.50 ng/ml are 777.8, 916.9, 1115.3, 1313.4, and 1780.3, respectively; highest sensitivity for a cortisol concentration of 4.50 ng/ml. A much higher maximum FOM* value indicates ultrasensitive detection of cortisol concentration with high accuracy. In addition to maximum FOM*, the calculated LOD for different cortisol concentrations of 0.72, 1.80, 3.60, and 4.50 ng/ml are 0.0038, 0.0067, 0.0554, and 0.0262, respectively, and plotted in Fig. 7.

Table 3 depicts all the performance parameters evaluated at considered cortisol concentrations in Figs (3–7). The best performance parameters obtained are sensitivity (343.78 $^{o}$/RIU), FOM (243.82 $RIU^{-1}$), maximum FOM* (1780.30), and finest LOD (0.026 ng/ml) for the proposed sensor.

Table 3. Performance parameters at different cortisol concentrations

| Cortisol Conc. (ng/ml) | $\theta_{SPR}$ (°) | FWHM (°) | S (°/RIU) | FOM (1/RIU) | LOD (ng/ml) | Maximum FOM* |
|---|---|---|---|---|---|---|
| 0.36 | 78.46 | 1.44 | - | - | - | |
| 0.72 | 78.56 | 1.44 | 314.67 | 218.52 | 0.0038 | 916.9 |
| 1.80 | 78.72 | 1.42 | 320.86 | 225.96 | 0.0067 | 1115.3 |
| 3.60 | 78.75 | 1.41 | 324.86 | 230.40 | 0.0554 | 1313.4 |
| 4.50 | 78.78 | 1.41 | 343.78 | 243.82 | 0.0262 | 1780.3 |

Furthermore, we also evaluated the normalized electric field component for the proposed SPR sensor using COMSOL simulation software. Fig. 8 illustrates the distribution of the normalized electric field ($E_z$) component of the evanescent field for the proposed sensor. According to Fig. 8(a), the maximum excitation of SPs occurs at the anisotropic BP/sensing layer interface and decays exponentially away from the interface [31]. Fig. 8 (b) represents the zoomed-in portion of the rectangular window marked in Fig. 8(a) to demonstrate the explicit representation of constituent layers of the proposed sensor. The normalized electric field component ($E_y$) normal to the interface for the proposed SPR sensor is also plotted in Fig. 9. It signifies the field distribution at each of the constituent layers' interfaces. The inset diagram within Fig. 9 demonstrates the peak normalized field intensity at the anisotropic BP/sensing layer interface, which decays exponentially away from the sensing interface. The penetration depth (PD) measures how deeply the field penetrates the sensing region and is defined as the distance covered by the electric field from its peak value to $1/e$ or 0.37% of its peaks value along normal to the interface [28]. The PD evaluated for the proposed sensor is 203 nm and is sufficient to detect the small size cortisol molecule [1].

Finally, we compared the proposed phosphorene-based cortisol sensor with significant cortisol sensors reported in the literature [11] [12] [13] [31]. As shown in Table 4, the proposed sensor has the highest sensitivity and finest LOD among all the reported sensors.

Table 4. Comparison of LOD for the proposed sensor with prominent cortisol sensor in literature.

| **Ref.** | **Authors (Year)** | **Sensor Type** | **S (°/ RIU)** | **LOD (ng/ml)** |
|---|---|---|---|---|
| [11] | Kämäräinen *et al.* (2018) | Electrochemical Immunosensor-based on direct competitive enzyme linked immunoassay | - | 0.6 |
| [12] | Dalirirad *et al.* (2019) | Au nanoparticles functionalized with Aptamer | - | 1.0 |
| [13] | Chen *et al.* (2016) | SPR Immunoassay (Denaturalized bovine-serum-albumin layer on SPR chip) | - | 1.0 |
| [31] | Cátia Leitão *et al.* (2021) | Plasmonic immunesensor-based on gold-coated tilted fiber Bragg grating (TFBG) | - | - |
| - | - | This work | 343.78 | 0.026 |

## 4 Conclusion

Relying on high adsorption energy, unique sensing ability, and selectivity of phosphorene for cortisol concentration and benefitting from tunable performance arising from the anisotropy of phosphorene, we reported tunable and sensitive cortisol detection using a surface plasmon resonance sensor configuration. A bimetallic combination of Cu (35 nm) and Ni (20 nm) is used to obtain a strong electromagnetic field in the analyte medium. The sensor performance is evaluated in terms of sensitivity, the figure of merit (FOM), the limit of detection (LOD), and a new parameter called FOM* and penetration depth (PD). Our proposed sensor has a very high sensitivity of 343.78 °/RIU, FOM of 243.82 $RIU^{-1}$, maximum FOM* of 1780.3, sharper LOD of 0.026 ng/ml, and a good PD of 203 nm. To the best of our knowledge, all these parameters are significantly higher than any of the existing state-of-the-art related to cortisol sensing. This research paves the way for the non-invasive, tunable detection of cortisol concentrations in saliva.

## Figure Caption List

**Fig.1** Proposed SPR sensor

**Fig. 2** Bi-Metal layer Thickness optimization curve

**Fig. 3** SPR curves for different cortisol concentrations in saliva

**Fig. 4** SPR angle, $R_{min.}$ and FWHM for different cortisol concentrations

**Fig. 5** Sensitivity and FOM for different cortisol concentrations

**Fig. 6** Angular figure of merit (FOM*) for different cortisol concentrations

**Fig. 7** LOD and Max. FOM* for different cortisol concentrations

**Fig. 8** Proposed sensor: (a) SPs field distribution (b) Inset plot

**Fig. 9** Normalized electric field ($E_y$) normal to the interface for proposed sensor

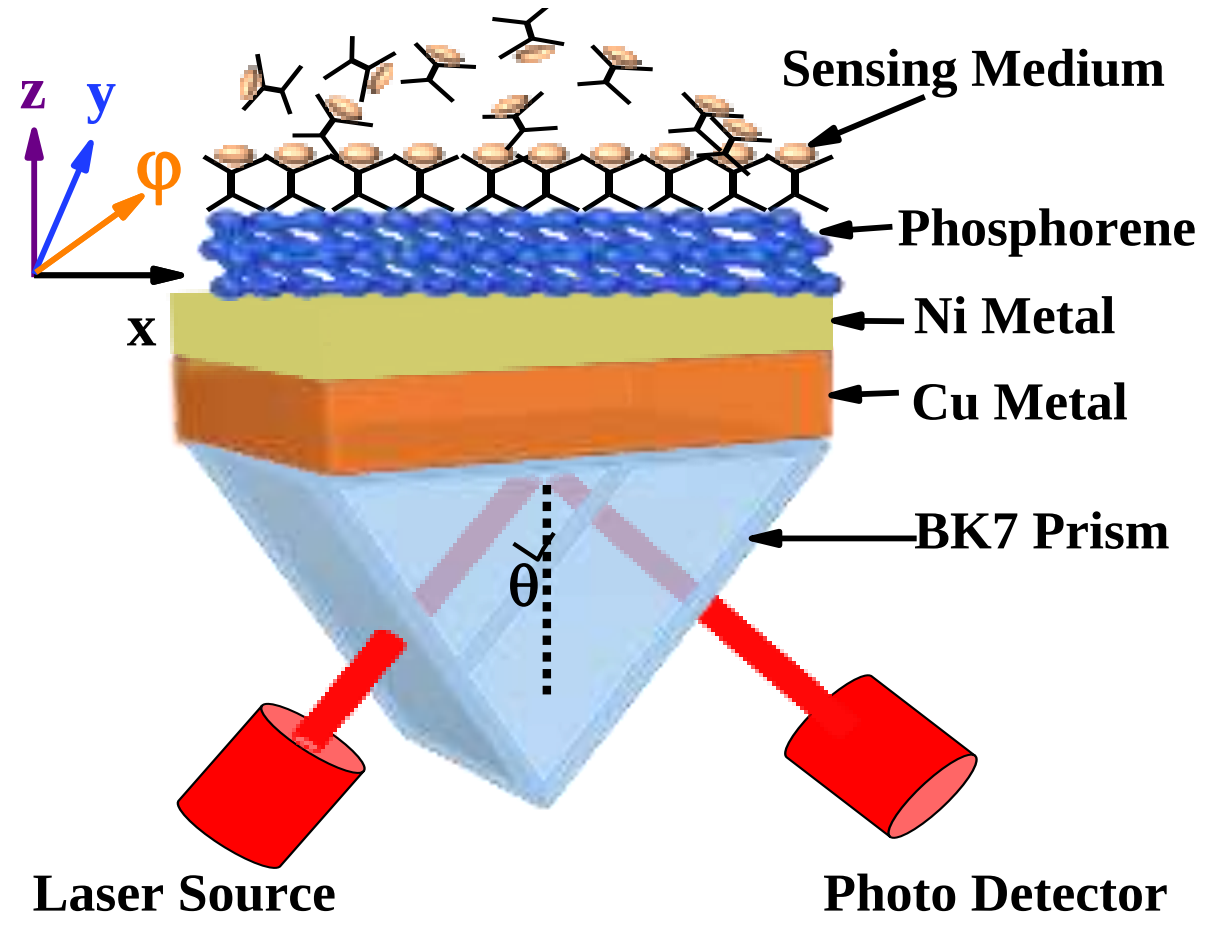

Fig. 1

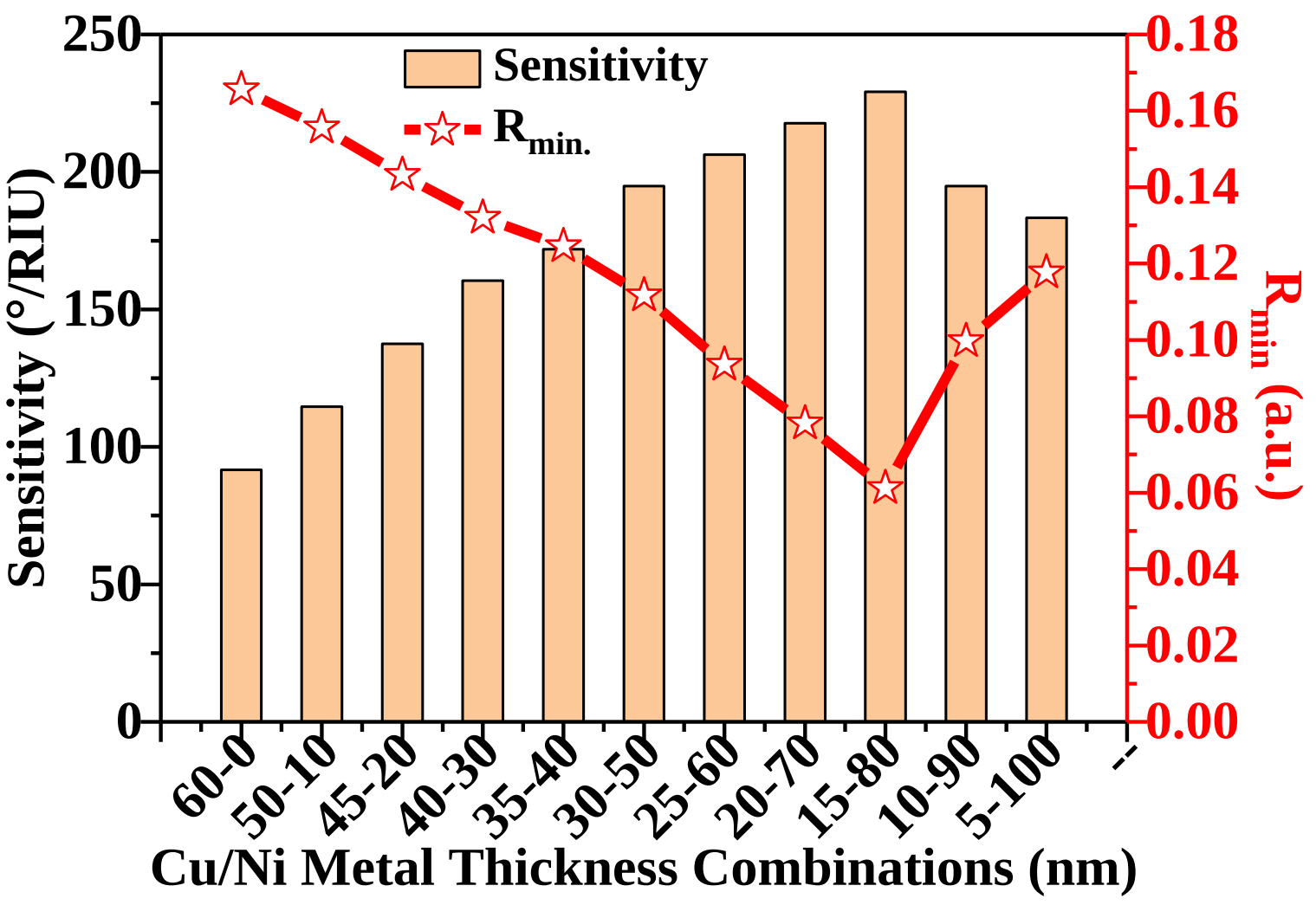

Fig. 2

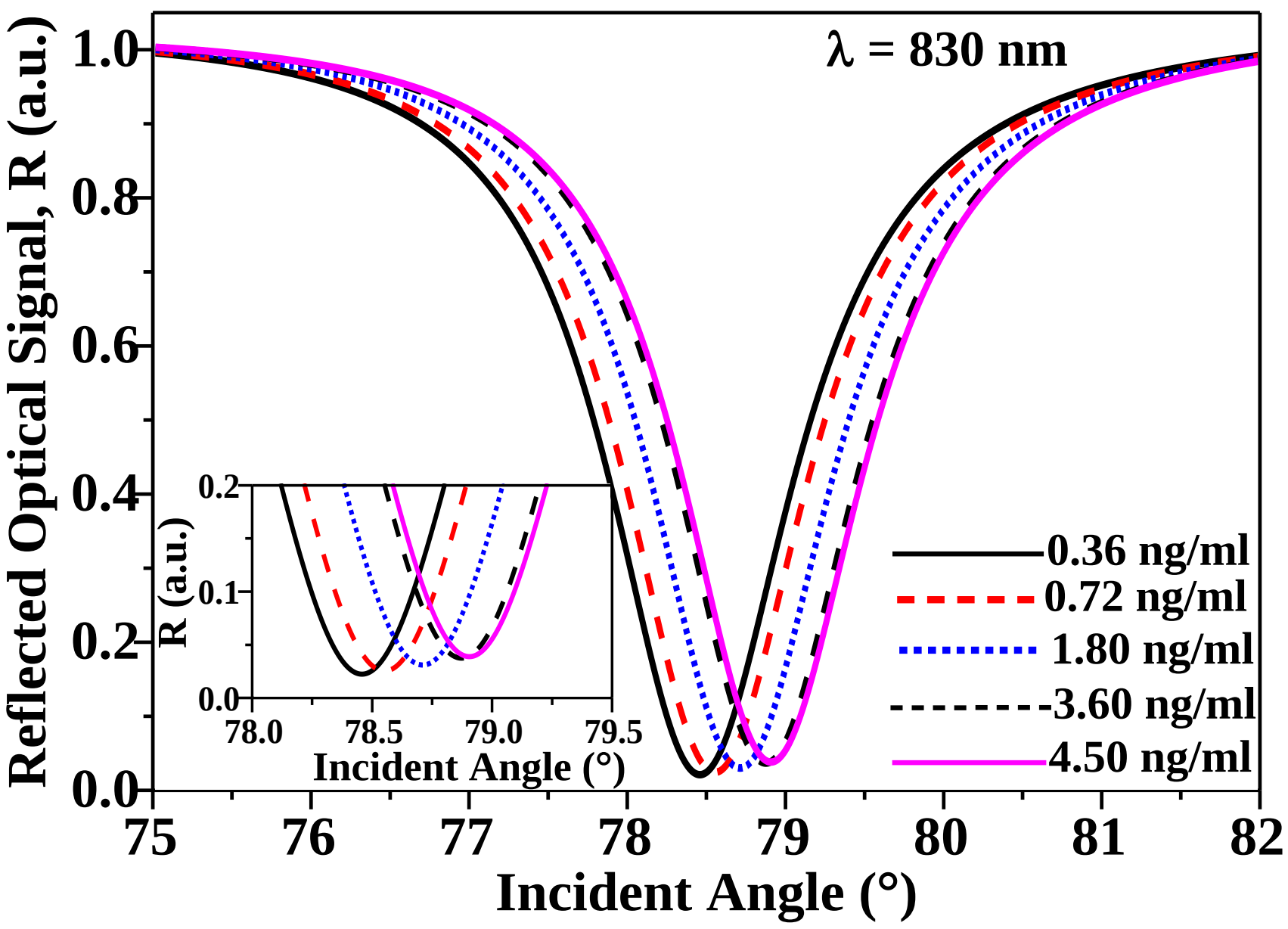

Fig. 3

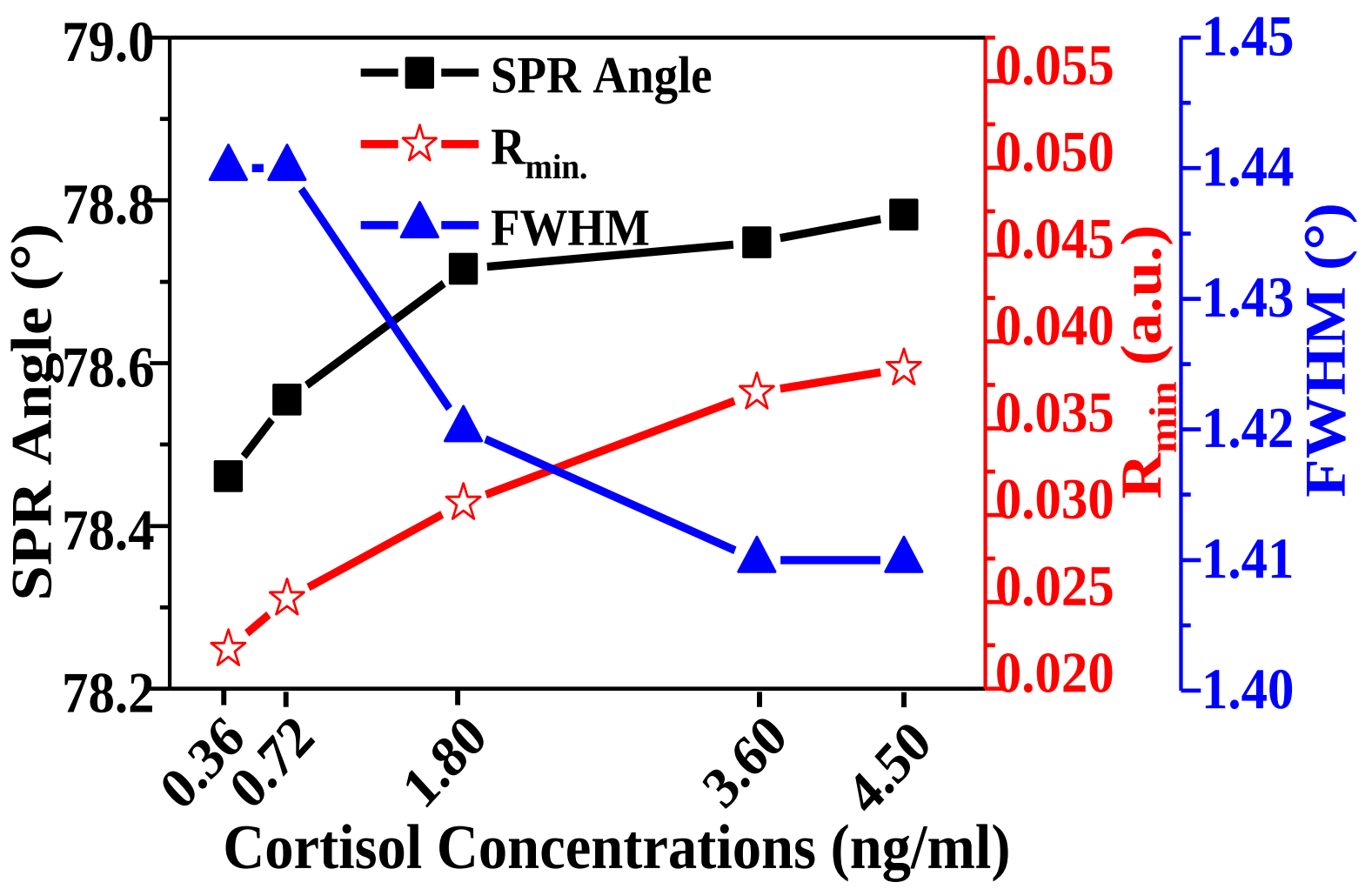

Fig. 4

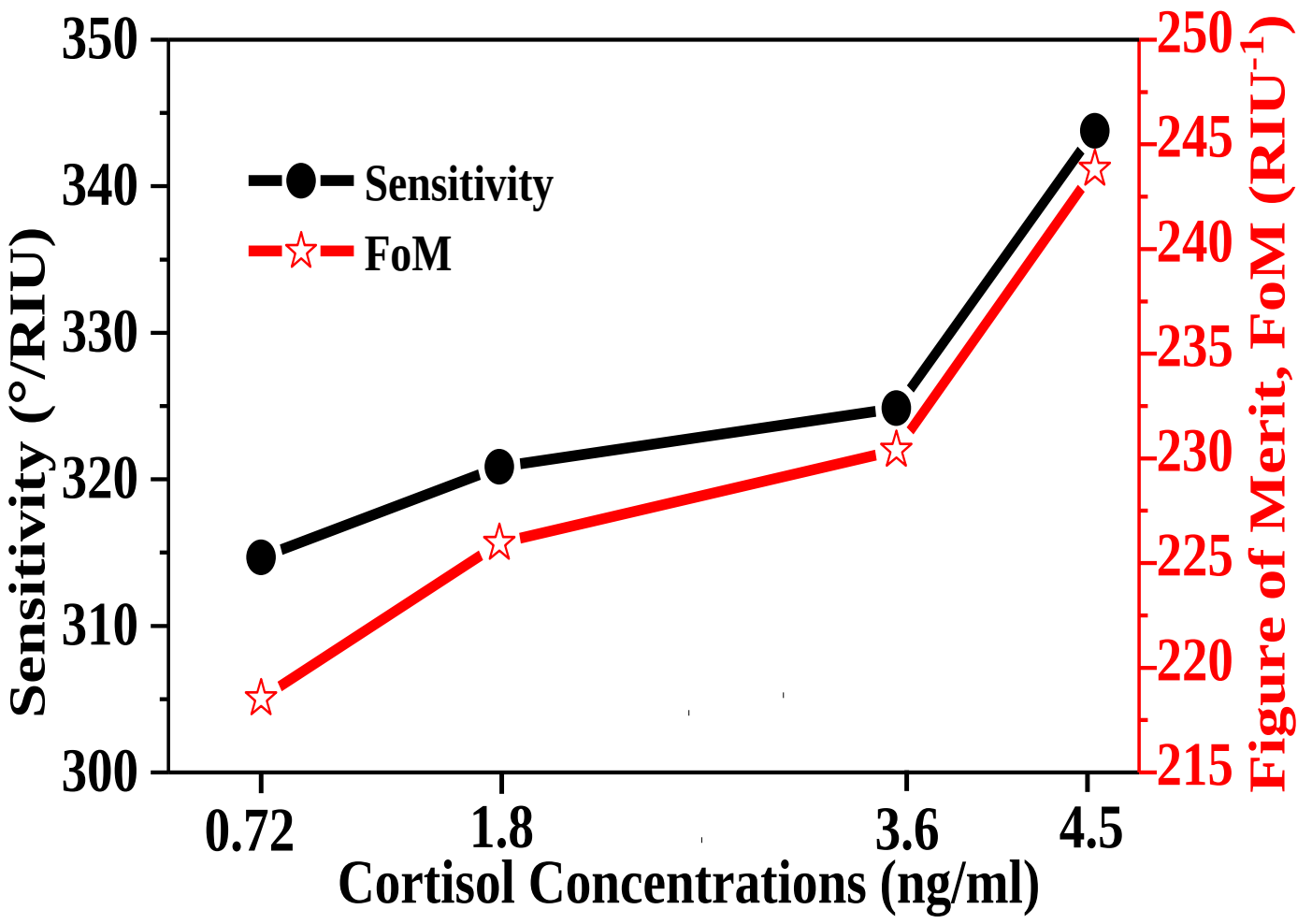

Fig. 5

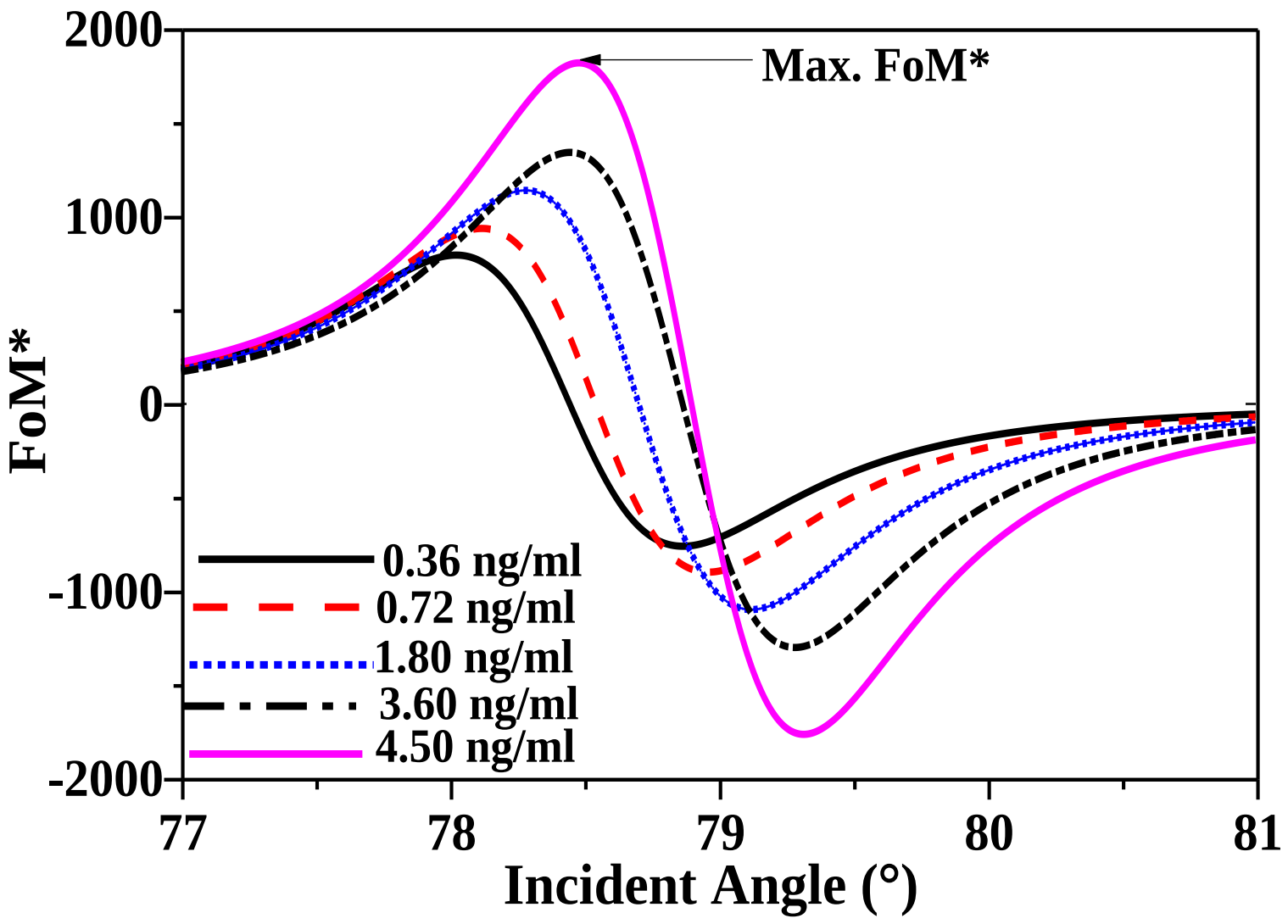

Fig. 6

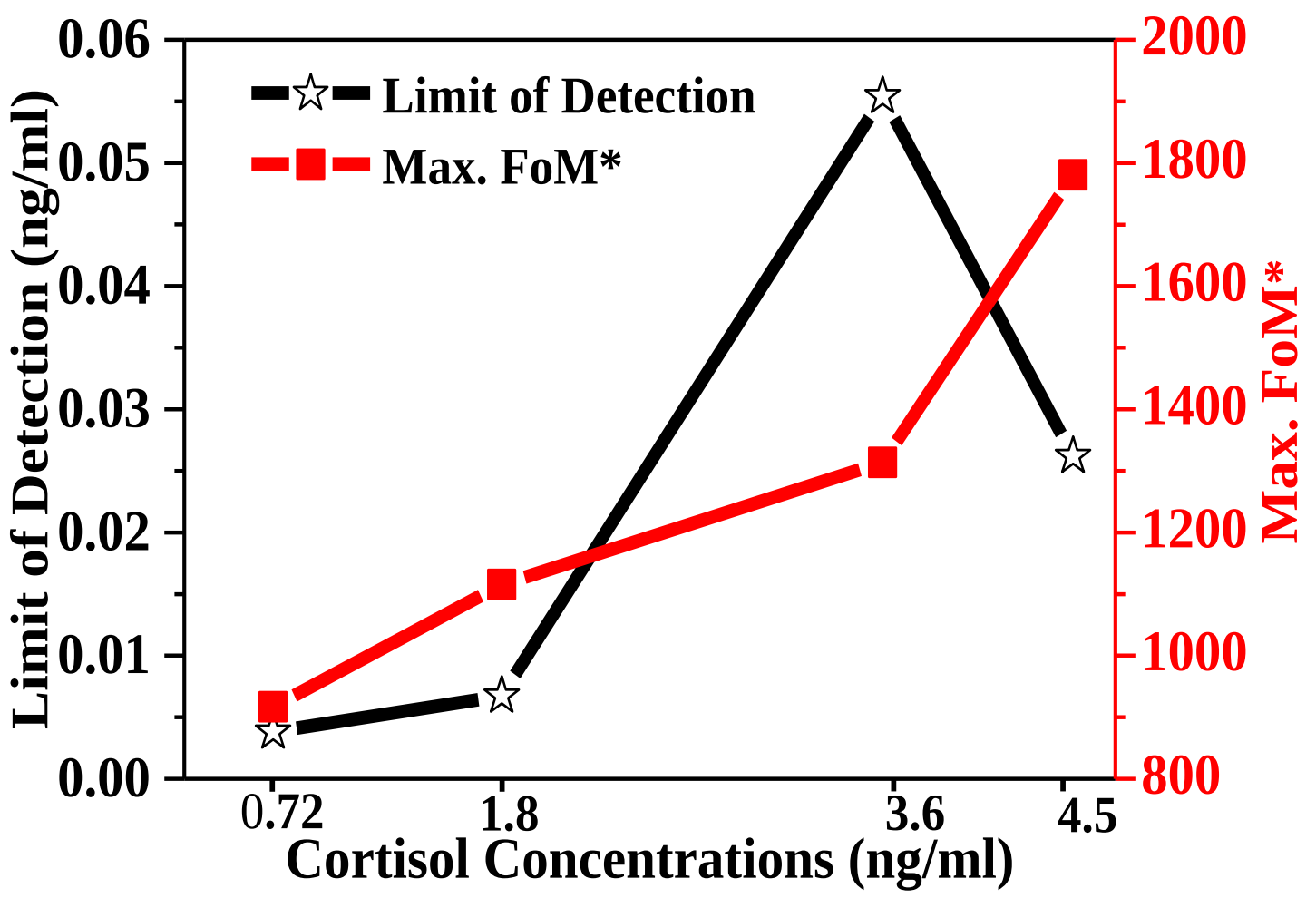

Fig. 7

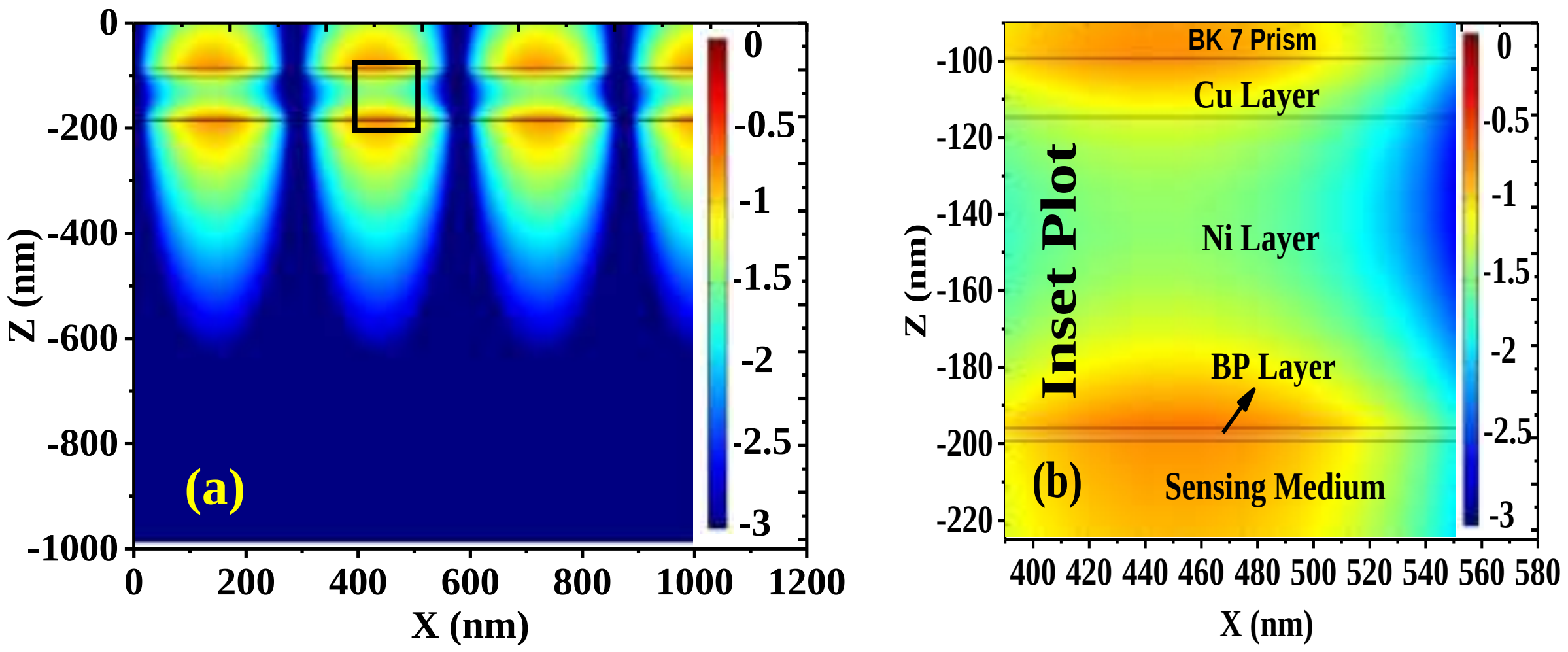

(a)
Z (nm)
X (nm)
0
-200
-400
-600
-800
-1000
0
200
400
600
800
1000
1200
0
-0.5
-1
-1.5
-2
-2.5
-3
(b)
Inset Plot
BK 7 Prism
Cu Layer
Ni Layer
BP Layer
Sensing Medium
Z (nm)
X (nm)
-100
-120
-140
-160
-180
-200
-220
400 420 440 460 480 500 520 540 560 580
0
-0.5
-1
-1.5
-2
-2.5
-3

Fig. 8

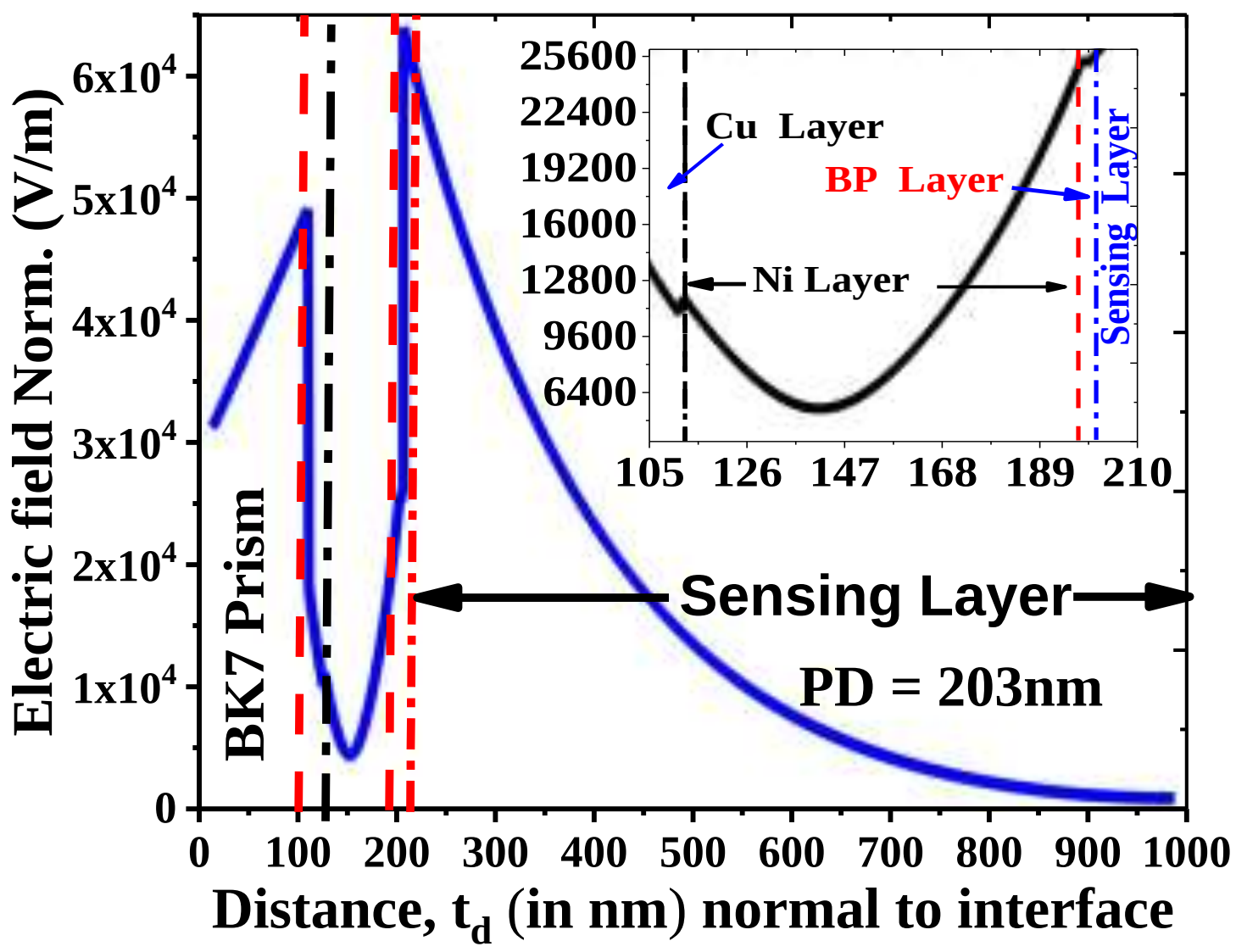
Electric field Norm. (V/m)
Distance, $t_d$ (in nm) normal to interface
BK7 Prism
Sensing Layer
PD = 203nm
Cu Layer
BP Layer
Ni Layer
Sensing Layer

Fig. 9